\documentclass[conference]{IEEEtran}
\usepackage{tabu}

\ifCLASSINFOpdf
   \usepackage[pdftex]{graphicx}
\else
\fi

\begin{document}
%
\title{Digital IP Protection Using Threshold Voltage Control}

\author{\IEEEauthorblockN{Joseph Davis, Niranjan Kulkarni, Jinghua Yang, Aykut Dengi, Sarma Vrudhula}
\IEEEauthorblockA{School of Computing, Informatics and Decision Systems Engineering\\
Arizona State University\\
Tempe, Arizona 85281\\}}

\maketitle

\begin{abstract}
This paper proposes a method to completely hide the functionality of a digital standard cell.  This is accomplished by a differential threshold logic gate (TLG).  A TLG with $n$ inputs implements a subset of Boolean functions of $n$ variables that are linear threshold functions. The output of such a gate is one if and only if an integer weighted linear arithmetic sum of the inputs equals or exceeds a given integer threshold.  We present a novel architecture of a TLG that not only allows a single TLG to implement a large number of complex logic functions, which would require multiple levels of logic when implemented using conventional logic primitives, but also allows the selection of that subset of functions by assignment of the transistor threshold voltages to the input transistors.  To obfuscate the functionality of the TLG, weights of some inputs are set to zero by setting their device threshold to be a high $V_t$.  The threshold voltage of the remaining transistors is set to low $V_t$ to increase their transconductance. The number of low $V_t$ transistors whose gates are driven by a given input $x_i$ determines the weight of that input.  The function of a TLG is not determined by the cell itself but rather the signals that are connected to its inputs. This makes it possible to hide the support set of the function by essentially removing some variable from the support set of the function.  This is done by selective assignment of high and low $V_t$ to the input transistors.  We describe how a standard cell library of TLGs can be mixed with conventional standard cells to realize complex logic circuits, whose function can never be discovered by reverse engineering.  A 32-bit Wallace tree multiplier and a 28-bit 4-tap filter were synthesized on an ST 65nm process, placed and routed, then simulated including extracted parastics with and without obfuscation.  By obfuscating the cells the delay was shown to increase by approximately 5\% at the cell level.  Both obfuscated designs had much lower area (25\%)  lower area and much lower dynamic power (30\%) than their  nonobfuscated CMOS counterparts, operating at the same  frequency.
\end{abstract}

%
\IEEEpeerreviewmaketitle

\section{Introduction}
With growing competition in the silicon industry, there is an increasing need to conceal the functionality of a circuit design.  This is due to the increasing use of reverse engineering by competitors, both foreign and domestic, aimed at stealing hardware intellectual property (IP).  Reverse engineering of a product can lead to companies loosing competitive advantages and losing market share to counterfeit products.  It has been estimated that counterfeit products cost the semiconductor industry \$4 billion a year \cite{SEMI}.  Examples of counterfeit products in such systems include airplane landing lights, radiation detectors, automotive breaking systems and defibrillators \cite{counterfeitWhitepaper}.  The consequences of using counterfeit chips include the premature failure of critical systems.  This is due to counterfeit chips that are manufactured cheaper and at a lower quality than the original.  Because of this, there is a need for innovative solutions to protect hardware IP from reverse engineering.  To combat these counterfeit parts many industries have adopted standards of manufacturing to minimize counterfeiting \cite{counterfeitSupplyChain}.

\subsection{Reverse Engineering}

By reverse engineering an integrated circuit (IC) counterfeiters can gain insight into the design of the circuit.  Reverse engineering itself has become a major challenge for counterfeiters due to the high cost and time required to reverse engineer sub-micron circuits.  One can reverse engineer an IC by learning the circuitry of the chip through optical imaging or the use of a scanning electron microscope (SEM) \cite{SEM_software}. In this process of each layer of the chip is removed one at a time and a picture is taken of that layer.  Due to the variety of different feature sizes in modern ICs, many different techniques are used today to remove each layer.  These techniques include dry etching, wet etching and polishing of the chip.  Each image will show the electrical connections in that layer.  By assembling the images, it is possible to reveal all electrical connections that constitute the chip.  There is commercial software available that can aid in the reconstruction of the electrical connections from SEM images.  This software can export spice level netlists for simulation of the IC \cite{SEM_software}.

To combat reverse engineering, researchers have come up with many different circuit level defenses.  These defenses can be separated into two different types: non-obfuscated and obfuscated defenses.  A non-obfuscated defense is an approach where an attacker can still see the functionality of the chip.  However, if they did reproduce the chip it could be detected as a counterfeit.  An obfuscated defense is where an attacker cannot learn the functionality of the chip.  The approach presented in this paper is an obfuscated defense.

\subsection{Non-Obfuscated Defenses}

A non-obfuscated defense can be viewed as a watermark or fingerprint, where a unique structure is added to the hardware to prove the identity of the intellectual property (IP) \cite{watermark}.  The difference between the two is a watermark is added to identify the designer and a fingerprint is added to identify the buyer \cite{Primer}.  Using this approach an attacker may be able to see the functionality of the chip, but may not be able to duplicate it.  This watermark or fingerprint can include a hidden feature in the datapath, control path, or a hidden layout feature that is hard to duplicate.  One solution is to add sheets of M1 to the filler cells.  By doing so, the cells become highly reflective when the chip is illuminated \cite{watermark}.  As this may be a detail that counterfeiters overlook, the reflectivity of the chip can be used to determine counterfeiting and detection of trojans.  Any extra hardware added by the fab will cause the surface to be less reflective, as there will be fewer filler cells.  Another watermarking approach is to include a unique number of clock buffers in the synthesis of the chip \cite{bufferWatermark}.  In doing so, if someone reverse engineers the chip and attempts to make a counterfeit, the number of buffers could be different.  This is because clock buffers are automatically added by the synthesis tool.  Therefore, the number of clock buffers can be used to identify a counterfeit chip.  As both of these methods are passive, there is no additional area or power cost for watermarking or fingerprinting.

A more advanced watermarking, or fingerprinting strategy is to add some functionality that is physically uncloneable.  These circuits are called physically uncloneable functions (PUF).  They work by creating a circuit where the functionality is highly dependent upon process variations and is therefore very hard to reproduce.  If the circuit is reverse engineered and cloned, it will not work due to differences in the process variations.  An obvious candidate for a PUF is a ring of an even number of inverters, as it will settle at a 1 or a 0 depending upon process variations, if it is not programmed~\cite{PUFRosc}.  The output is measured, the circuit is reset and the output is measured again.  A golden distribution of outputs will be generated.  When, an unverified chip is measured, the distribution of outputs should be similar.  If it is not, than the circuit is a counterfeit.  This is because the output of an uninitialized memory element is a function of the parasitics.  So as long as the parasitics are the same from device to device, the distribution of outputs for the uninitialized memory element should be the same.  Therefore, an attacker may be able to counterfeit the chip, but the PUF will act as a watermark.

\subsection{Obfuscated Defenses}

The other class of circuit level defenses is obfuscated defenses, in which an attacker may not be able to reverse engineer the functionality of the chip.  One approach of this type is to make the layouts of all of the standard cells the same, so that a NOR gate looks identical to a NAND gate.  This is done through hiding the connections in between transistors.  This can be done by using false vias between layers~\cite{vias} to hide connections.  The disadvantage of this approach is that it requires a special manufacturing process for the false vias and increases the capacitance among wires.  This approach substantially increases the power, delay and area of the cell library by approximately 500\%, 150\% and 400\% respectively.  Because of this, the authors propose only obfuscating 5\% of the total gates in a large design.  The total circuit overhead was approximately 80\% increase in power, 70\% increase in area and 50\% increase in delay for the ISCAS test circuits.

Another method to conceal connections is through adding connections through diffusion \cite{diffusion}.  By doing so, one can create inverters that are stuck at a logic one, or a logic zero.  Using these obfuscated inverters, one can hide the true number of inputs into a function.  The disadvantage of this approach is a large increase in area and delay.  One paper reported that this approach increased the area by 587\% and the delay by 250\% \cite{diffusion}.

Other non-hardware defenses include the use of \textit{software keys} to define the datapath flow \cite{softwareKey}.  A software key is a control signal that changes the control flow of a datapath.  The software key can be a selector bit for a Mux \cite{softwareKey}, or a lookup table \cite{swkeyLUT}, or an XOR gate \cite{swkeyXor}.  This approach increases the area by an average of 0.63\% and the power by 2.6\% \cite{softwareKey}.  The overhead was so low because the authors only obfuscated 5\% of the total number of cells in the design.

One of the most secure ways to obfuscate a function is by selectively changing the threshold voltage of certain transistors.  Since the threshold voltage depends upon the gate oxide thickness, it is very difficult to detect changes in $V_t$ through SEM images.  This approach has been used previously to obfuscate standard cells at the expense of static DC current~\cite{VtControl}.  In~\cite{VtControl},  the logical inputs are made to a pMOS pull up network.  The threshold voltage of the nMOS pull down network is varied to create either a NAND, or a NOR gate.  In this paper we expand on this idea to obfuscate larger logical functions, by varying $V_t$ and without creating DC currents or degrading performance or area or power. 

The remainder of this paper is divided up into the following sections.  Section II gives a brief introduction to threshold logic and the circuit level implementation of a threshold logic gate (TLG).  Section III shows cell level obfuscation and compares obfuscated and nonobfuscated threshold gates to their CMOS counterparts.  The functions used for comparison are a 3-input AND function and a 3-input XOR function.  Section IV uses the obfuscated cell library proposed in Section III to build larger circuits.  An obfuscated Wallace Tree Multiplier and an obfuscated FIR Filter are synthesized and placed and routed.  The circuits are compared to their nonobfuscated CMOS counterparts for comparison of area and power.  Section V draws conclusions on the obfuscated cells and wraps up the paper.

\section{Threshold Logic}
Threshold logic is a subset of Boolean logic.  A Boolean logic function is defined as a threshold logic function if the function $\textit{f}(x_1,x_2,...,x_n)$ along with its associated weights $(w_1,w_2,...)$ and threshold $T$ satisfy

\begin{eqnarray}
\label{eq_threshold_fn_def}
f(x_1, x_2, \cdots, x_n) & = & \left\{ \begin{array}{ll}
1 & \mbox{if} ~\sum_{i=i}^n w_i x_i \geq T \\
0 & \mbox{otherwise}.
\end{array}
\right.
\end{eqnarray}

Where the weights $w_n$, the threshold $T$ are integers and the inputs $x_n$ are Boolean variables.  A threshold function will be represented as $[w_1,w_2,...w_n;T]$.  Boolean threshold functions can be determined through Binary Decision Diagrams \cite{BDD}.  Circuit implementation of threshold logic has been the focus of much research \cite{ThresholdLogic}.  It has been previously shown that by implementing logic with threshold cells, one can save area and power, compared to their CMOS counterpart.  

\subsection{Circuit Implementation of Threshold Logic}

\begin{figure}[!t]
\centering
\includegraphics[scale=0.32]{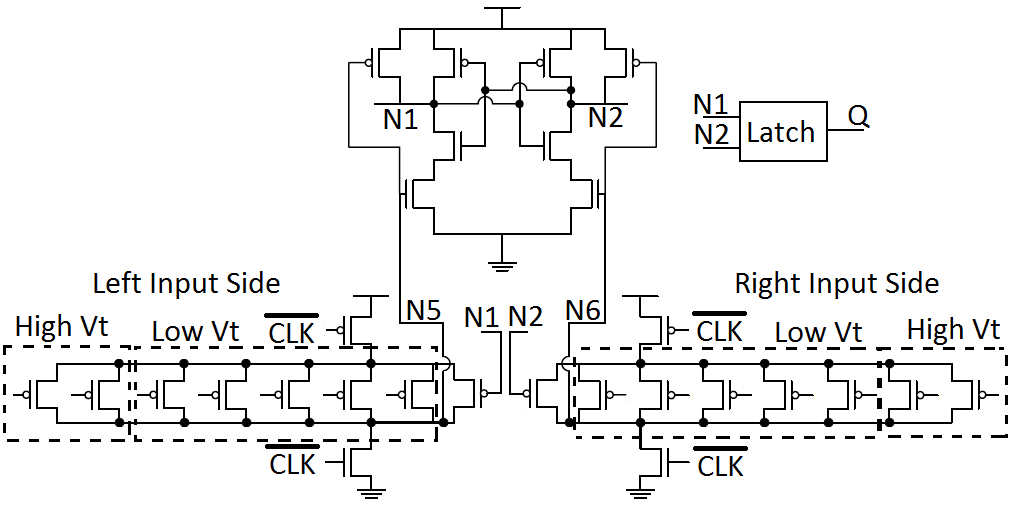}
\caption{TLG With Obfuscated Inputs}
\label{fig:PNAND7}
\end{figure}

A circuit that evaluates Equation \ref{eq_threshold_fn_def} this paper will refer to as a threshold logic gate (TLG).  A TLG will make a comparison between $n$ number of inputs and their associated weights $w$ and a threshold $T$.  To identify different threshold cells, TLGs will follow the naming scheme TLG-n, where n is the number of inputs.  In the proposed TLG, shown in Figure \ref{fig:PNAND7}, two input networks act as the logical inputs and a threshold value $T$.  Assignment of signals to the input network determine the functionality of the cell.  For example, the equation $f = a \vee bc$ is represented as $f = [2,1,1;2]$.  To avoid any case where the input side is equal to the threshold side we double the input variables and increase the threshold by 50\%.  Therefore, the equation becomes $f = [4,2,2;3]$.  This produces a signal assignment of left input network = $\bar{a},\bar{a},\bar{a},\bar{a},\bar{b},\bar{b},\bar{c},\bar{c}$ and right input network = $0,0,0$.  Note that the signals are inverted due to the inputs being pMOS in Figure \ref{fig:PNAND7}.  This signal assignment is unoptimized as it requires 8 transistors on one side and 3 transistors on the other side.  Some of the signals are moved from one side to the other for an optimal signal assignment of $f = a \vee bc$ to be left input network= $\bar{a},\bar{a},\bar{b},\bar{c},\bar{c}$ right input network = $a,a,b,1,1$.  Note that all of these inputs will be connected to low $V_t$ transistors, as they are unobfuscated inputs.  Random signals will be connected to high $V_t$ inputs to hide the functionality of the cell.  In the above example a TLG-5 with 1 or 2 extra obfuscated inputs can be used.

The TLG shown in Figure \ref{fig:PNAND7} consists of three parts: an input network, a sense amplifier and a latch.  The circuit works in the following manor:
\begin{enumerate}
\item The logical inputs and the threshold are applied to the left and right input networks. 
\item The clock rises from a 0 to a 1, causing nodes N5 and N6 to rise from a 0 to a 1.  The rate at which they rise is dependent upon the number of transistors on and the associated transconductance (gm) of each transistor. 
\item The race condition between N5 and N6 will set the sense amplifier to either a 1 or 0.
\item The sense amplifier will set the latch and the latched value will appear as output $Q$. 
\item The clock will fall from 1 to 0 causing N5 and N6 to be reset to 0, while maintaining the output $Q$ in the latch.
\end{enumerate} 
Therefore, the circuit acts as an edge triggered flip flop.  Large Boolean functions can be \textit{absorbed} by the cell, as the functionality of the cell is determined by the input signal assignment.  Provided that the absorbed logic has a delay of one clock cycle.  Therefore, it requires lower energy, area and delay compared to a CMOS implementation of the same Boolean function \cite{ThresholdLogic}.

\subsection{Obfuscation of TLG Circuit}
The TLG topology shown in Figure \ref{fig:PNAND7} can be easily obfuscated.  As the functionality of the TLG is determined by the input selection, it is a natural candidate for an obfuscated cell library.  To obfuscate the functionality of the TLG gate this paper recommends increasing the $V_t$ on some input transistors and lowering the $V_t$ on other input transistors.  By doing so, the low $V_t$ transistors will have a large transconductance, therefore, a large weight in Equation \ref{eq_threshold_fn_def}; high $V_t$ transistors will have a low transconductance, therefore a small weight in Equation \ref{eq_threshold_fn_def}.  If the difference in $V_t$ is large enough, than the difference in weights will be large enough to effectively remove the small weights for high $V_t$ transistors.  So by setting two of the inputs on each side to high $V_t$ and setting the rest of the inputs to low $V_t$, the high $V_t$ inputs will be logically removed from the calculation of Equation \ref{eq_threshold_fn_def}.  Also, the weight on the other inputs will only be determined by the number of transistors that are connected to the input signal.  Therefore, if someone were to reverse engineer the circuit they would not observe that some of the inputs are logically separated from the calculation of the function without measuring the $V_t$ of every input transistor.  Since it is very difficult to observe $V_t$ in a large layout, this approach is difficult to defeat.  The novelty of this approach is that it means an attacker would not even know that the circuit was obfuscated.  This defense can also hide the functionality of the circuit from the foundry.  As the $V_t$ mask is used to determine obfuscated and nonobfuscated inputs.  Without it, it is impossible to know the operation of the circuit.  Because of this, one could split the fabrication up into two different foundries.  One foundry would only apply the $V_t$ mask and the other foundry would fabricate the rest of the chip.  Another advantage of the TLG architecture is since it is comparator based, the dynamic power consumed each clock cycle is approximately the same regardless of what function is being calculated.  Therefore, the circuit would be immune to power attacks.

\begin{table}[]
\centering
\caption{Summary of Obfuscated Cell Library}
\label{my-label}
\begin{tabular}{|c|c|c|c|c|}
\hline
\begin{tabular}[c]{@{}c@{}}Name\\ \# of Obfuscated Devices on\\ One Side\end{tabular}  & \begin{tabular}[c]{@{}c@{}}TLG-3\\ 1\end{tabular} & \begin{tabular}[c]{@{}c@{}}TLG-3\\ 2\end{tabular} & \begin{tabular}[c]{@{}c@{}}TLG-5\\ 1\end{tabular} & \begin{tabular}[c]{@{}c@{}}TLG-5\\ 2\end{tabular} \\ \hline
\begin{tabular}[c]{@{}c@{}}Combinations of\\ Obfuscated Functions\end{tabular}         & 9                                                 & 9                                                 & 25                                                & 100                                               \\ \hline
\begin{tabular}[c]{@{}c@{}}Name\\ \# of Obfuscated Devices on\\  One Side\end{tabular} & \begin{tabular}[c]{@{}c@{}}TLG-7\\ 1\end{tabular} & \begin{tabular}[c]{@{}c@{}}TLG-7\\ 2\end{tabular} & \begin{tabular}[c]{@{}c@{}}TLG-9\\ 1\end{tabular} &                                                   \\ \hline
\begin{tabular}[c]{@{}c@{}}Combinations of\\ Obfuscated Functions\end{tabular}         & 49                                                & 441                                               & 81                                                &                                                   \\ \hline
\end{tabular}
\label{tab:lib_summary}
\end{table}

Using this technique a cell library of 7 cells was designed and laid out in 65nm technology.  The cells were a TLG-3, TLG-5, TLG-7 with 1 and 2 extra obfuscated inputs and a TLG-9 with one extra obfuscated input.  A summary of how many obfuscated functions each cell can produce is shown in Table \ref{tab:lib_summary}.  This number is calculated by 

\begin{eqnarray}
\label{obf_fn_number}
Y = {(n+k)\choose k}^2 .
\end{eqnarray}

Where $Y$ is the number of different ways the gate can be obfuscated, $n$ in the number of valid inputs, and $k$ is the number of obfuscated inputs.  Since threshold gates are high fan-in gates, more than one input may be obfuscated.  This means a single obfuscated threshold gate can represent more functions than a single obfuscated CMOS gate.  For example, an obfuscated CMOS cell in \cite{vias} can represent one of three separate functions, while a TLG-7 with 2 extra obfuscated inputs can be obfuscated in 1296 different ways.  These cells were characterized and used in high level synthesis to obfuscate larger circuits.  To maximize the yield of an obfuscated TLG, the supply voltage was lowered to 1.1V.  In doing so, the $V_{gs}$ of the input transistors are lowered, allowing the transistors to be more sensitive to $V_t$ variations.  Lowering the supply voltage lower than 1.1V causes the circuit to fail more often because of errors in the sense amplifier.  The yield of the circuit will be affected by power supply noise, as will every comparator based circuit.  




A distribution of the $V_t$ for pMOS transistors in 65nm technology is shown in Table \ref{tab:Vts}.  In the TLG-9 circuit in Figure \ref{fig:PNAND7}, these different $V_t$ correspond to peak drain currents on the input transistors of 15.3uA for the low $V_t$ and 4.54uA for the high $V_t$.  So the drain current of the low $V_t$ transistor is 3.37 times larger than the drain current for the high $V_t$ transistor.  In Figure \ref{fig:PNAND7} since two inputs are obfuscated away, the circuit would be referred to as a TLG-7 with two obfuscated inputs.  Since this circuit works by comparing two rising voltages on nodes N5 and N6, the difference between the two nodes determines the robustness of the cell.  Therefore, a process with a large difference in $V_t$s is required for this technique to work.  If the difference in $V_t$s is too small then the logically removed inputs will begin to have an effect on the circuit and the circuit will not be obfuscated.

\begin{table}[!t]
\centering
\caption{$V_t$ distribution of pMOS for 10000 samples at 1.1V 25C}
\begin{tabular}{|c|c|c|} \hline
&Mean&Standard Deviation\\ \hline
LVT&-280mV&19.5mV\\ \hline
SVT&-433mV&22.8mV\\ \hline
HVT&-604mV&21.7mV\\ 
\hline\end{tabular}
\label{tab:Vts}
\end{table}

\section{Cell Level Obfuscation}
To prove the cell level operation of the obfuscated TLG two different functions were tested with an obfuscated TLG cell.  The two cell level designs demonstrated in this paper are an obfuscated 3-input AND cell and an obfuscated 3-input XOR cell.  For each circuit we compare the functionally sequential equivalent CMOS version to our TLG version, both nonobfuscated and obfuscated.  The TLG version is termed Hybrid to differentiate it from CMOS.  Experiments were performed on extracted layout netlists using HSpice.  The setup time, nominal clock to Q, and total delay for the obfuscated and non-obfuscated versions of each circuit was calculated.  The setup time was defined as the input to clock delay that increased the nominal clock to Q delay by 10\%.  The total delay was defined as the sum of the setup time and the nominal clock to Q delay.

\subsection{Obfuscated 3-Input AND Cell}
A 3-input AND function can be constructed with a single TLG-5.  The threshold function representing an AND gate is $f(a,b,c)=[1,1,1;3]$.  This corresponds to an input signal assignment of left input network = $\bar{a}, 1, 1, 1, 1$ and right input network =$a, b, b, c, c$.  The inputs set to 1 are removed from equation, as the input transistors are pMOS.  To obfuscate the function a TLG-5 with two obfuscated inputs was used to implement the function.  The unobfuscated inputs were connected to the low $V_t$ transistors and the obfuscated inputs were connected to the high $V_t$ transistors.  The obfuscated cell was compared to an unobfuscated TLG-5.  The setup time and clock to Q delay of the CMOS circuit, nonobfuscated and unobfuscated versions of the cell are shown in Table \ref{tab:AND} for layout simulation the slow corner of 1.1V 105C SS, with a load of 20fF.  Because an AND function is combinational, yet the TLG is sequential, a D flip flop was connected to the output of the CMOS AND gate.  The table shows that the TLG version decreases the setup time and increases the clock to Q delay compared to the CMOS version.  This is due to the structure of the TLG cell.  By having the clock transistors part of the input network as seen in Figure \ref{fig:PNAND7} the setup time becomes really small.  Table \ref{tab:AND} shows the hybrid version of the obfuscated cells has roughly equal delay compared to the CMOS version.  The increase in delay between the two is 1.1\%.  Therefore, at the cell level this approach of obfuscation does not add much overhead in terms of delay.

\begin{table}[!t]
\centering
\caption{Comparison of CMOS, Hybrid Obfuscated and Hybrid Nonobfuscated 3-Input AND Gate, 1.1V 105C SS with 20fF load}
\begin{tabular}{|c|c|c|c|} \hline
& CMOS& Hybrid NonObfuscated& Hybrid Obfuscated\\ \hline
Setup Time&186ps&-58ps&8ps\\ \hline
Clock to Q&258ps&353ps&441ps\\ \hline
Total Delay&444ps&295ps&449ps\\ 
\hline\end{tabular}
\label{tab:AND}
\end{table}

\subsection{Obfuscated 3-Input XOR Cell}
\begin{figure}[!t]
\centering
\includegraphics[scale=0.23]{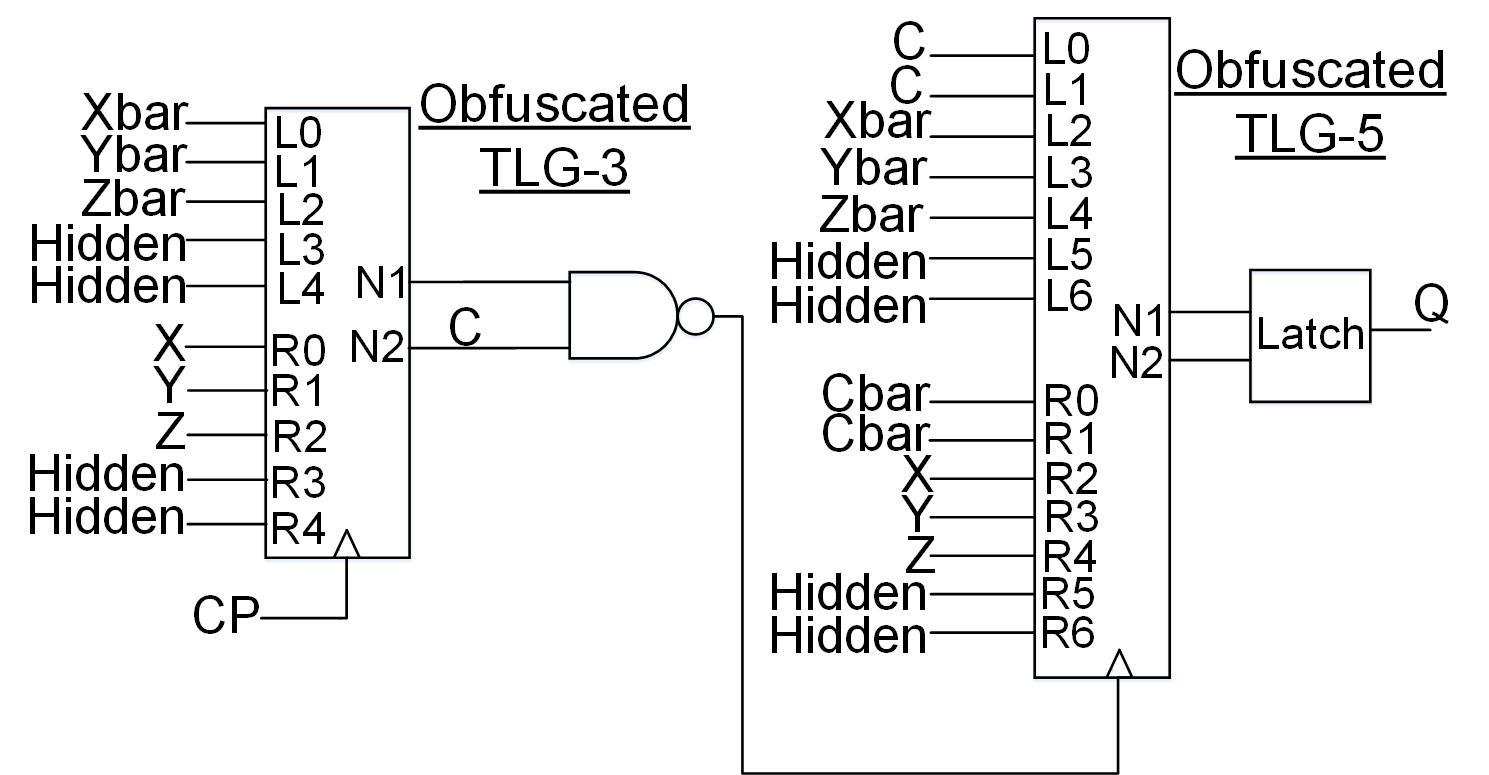}
\caption{3-Input XOR Gate with Obfuscated TLG-7}
\label{fig:XOR}
\end{figure}

To demonstrate that this approach for larger circuits a 3-input XOR cell was demonstrated.  Because the truth table of the XOR function is not monotonic, it is not a threshold function.  Instead, it is calculated using two levels of threshold gates.  The XOR cell was formed using an obstructed TLG-3 and an obfuscated TLG-5 in 65nm.  It consists of a TLG-3 with outputs N1 and N2 being fed into a NAND gate, which functions as the clock signal for a TLG-5.  In doing so the XOR function is calculated in a single clock cycle.  The complete circuit is show in Figure \ref{fig:XOR}.  In the figure the Ls and Rs represent the left and right input network of a TLG.  Input transistors L3, L4, R3 and R4 were set to high $V_t$ on the TLG-3 and input transistors L5, L6, R5 and R6 were set to high $V_t$ on the TLG-5.  The rest of the input transistors were set to low $V_t$.  In doing so, these eight inputs were logically removed from the calculation of the XOR function.  These fake inputs can be chosen at random by the circuit designer to hide the cell's function.  This circuit was compared against an unobfuscated version of itself in post-layout simulations.  Again, D flip flops were added to the output of the CMOS XOR gate for a direct comparison between CMOS and TLG versions.  At the corner of 1.1V 105C SS and a load of 20fF, the delay of the obfuscated and unobfuscated TLG cells along with their CMOS counterpart are shown in Table \ref{tab:XOR}.  The table shows a 5.1\% increase in delay between CMOS and the obfuscated hybrid version.  This increase in delay can be attributed to a more complicated hybrid version of the XOR function, than a CMOS implementation.

\begin{table}[!t]
\centering
\caption{Comparison of CMOS, Hybrid Nonobfuscated and Hybrid Obfuscated 3-Input XOR Gate, 1.1V 105C SS with 20fF load}
\begin{tabular}{|c|c|c|c|} \hline
&CMOS& Hybrid Nonobfuscated& Hybrid Obfuscated\\ \hline
Setup Time&264ps&-61ps&-61ps\\ \hline
Clock to Q&258ps&608ps&611ps\\ \hline
Total Delay&522ps&547ps&550ps\\
\hline\end{tabular}
\label{tab:XOR}
\end{table}

\section{Circuit Level Obfuscation}
In this section we use the cell library to obfuscate a Wallace Tree Multiplier and a FIR Filter.  These circuits were synthesized first using nonobfuscated CMOS standard cells.  Next, the circuit was "hybridized."  The process of hybridization is described in detail by Kulkarni in \cite{ThresholdLogic}.  The process involves searching the synthesized netlist for threshold functions and then replacing that logic with a TLG cell.  The CMOS and hybrid versions of the circuit were compared for power, area and delay.  The design flow of hybridizing a circuit is as follows:

\begin{enumerate}
\item A standard cell library of obfuscated TLGs was created in 65nm technology.
\item The cell library was characterized for worst case delay, power and leakage.  For characterization the obfuscated inputs are connected to Vdd.
\item The Verilog description of a multiplier and of a filter were synthesized using Cadence RTL compiler.  A CMOS standard cell library was used in synthesis.  The corner of SS 105C 1.1V was chosen for synthesis.
\item The synthesized netlist produced was hybridized using the method in \cite{ThresholdLogic}.  That is, the RTL netlist was edited so some of the CMOS cells were replaced with TLG cells.  Signals connected to the obfuscated inputs were chosen at random from inputs in the combinational fan-in to the TLG cell.
\item The hybridized RTL netlist was synthesized again to meet a targeted clock period.  The TLG cells placed in the netlist were untouched during synthesis. 
\item Both the synthesized CMOS netlist and the synthesized hybrid netlist were placed and routed separately in Cadence Encounter.
\item The place and routed netlists for both the CMOS and hybrid designs were simulated with delay annotation in Verilog.  A large number of inputs corresponding to 30\% switching activity were applied to the two designs in Verilog.  The Verilog netlist, along with the extracted circuit libraries were given to Synopsys Primetime to calculate the dynamic power and leakage power of the different designs.
\end{enumerate}

\subsection{Obfuscating a Multiplier}
The multiplier was a 32-bit 2 stage signed Wallace Tree Multiplier.  A CMOS version and a hybrid version were constructed to compare the two.  They were simulated at 666MHz with a 30\% switching activity.  A comparison of the two versions is shown in Table \ref{tab:Mult}.  The table shows a decrease of approximately 27\% area, 30\% dynamic power and 45.2\% leakage power of the hybridized multiplier compared to the CMOS version.  This is due to the absorption of logic by the TLG cells.  It is also worth noting that the CMOS version is not obfuscated so it is easy for an attacker to reverse engineer the design.  However, the hybrid version of the multiplier had 112 obfuscated TLGs out of a total of 4382 cells in the design.  Since each obfuscated TLG cell can represent 2 or more functions and there are 112 obfuscated TLGs in the circuit, the circuit has potentially 2\textsuperscript{112} different functions it can represent.  Making the circuit difficult to duplicate.

\begin{table}[!t]
\centering
\caption{Comparison of Multiplier in CMOS and Obfuscated Hybrid at SS 1.1V 105C}
\begin{tabular}{|c|c|c|} \hline
& CMOS& Hybrid Obfuscated\\ \hline
Area&43680 um X um&31856 um X um\\ \hline
Dynamic Power& 31mW& 21.5mW\\ \hline
Leakage Power&25.5uW&13.96uW\\ 
\hline\end{tabular}
\label{tab:Mult}
\end{table}

\subsection{Obfuscating a FIR Filter}
A filter was a second circuit that we used to compare CMOS to hybrid designs.  The FIR Filter was a 32-bit 4 tap 2 stage filter. The circuit was simulated at a frequency of 454MHz with a 30\% switching activity at the inputs.  The two versions of the circuit are compared in Table \ref{tab:Filter} above.  The hybrid version is 25\% lower area, 30\% lower dynamic power and 44\% lower leakage power than the CMOS version.  This circuit, along with the multiplier show how logic is absorbed by the TLGs during hybridization.  The hybrid version of the filter consisted of 93 obfuscated TLGs out of a total of 13254 cells. Since each obfuscated TLG can consist of 2 or more functions, the 93 TLGs can correspond to 2\textsuperscript{93} different functions.  Again, this shows it is very difficult to replicate the circuit's functionality.  It consisted of TLG-5s, with both 1 and 2 obfuscated inputs and TLG-7s and TLG-9s both with 1 obfuscated input.  These obfuscated inputs were randomly connected to combinational inputs in the logic cone.

\begin{table}[!t]
\centering
\caption{Comparison of FIR Filter in CMOS and Obfuscated Hybrid at SS 1.1V 105C}
\begin{tabular}{|c|c|c|} \hline
& CMOS& Hybrid Obfuscated\\ \hline
Area&124605 um X um&92720 um X um\\ \hline
Dynamic Power& 58.2mW& 40.2mW\\ \hline
Leakage Power& 83.57uW& 46.49uW\\ 
\hline\end{tabular}
\label{tab:Filter}
\end{table}

\section{Conclusion}
This paper presents a new way to obfuscate the functionality of digital circuits and protect against counterfeiting.  The approach is to use different threshold voltages on input transistors in order to logically remove certain inputs from the functionality of the circuit.  This is possible in the TLG circuit, where the weight of an input is defined by the threshold voltage.  By increasing the threshold voltage of an input transistor, the associated weight of a certain input can be set to 0.  As it is very difficult to reverse engineer the threshold voltage of a transistor, an attacker cannot discover the functionality of the TLG circuit.  We have demonstrated this approach in 65nm extracted layout simulations for an AND gate and an XOR gate at the cell level.  The two gates showed that by obfuscating the circuit the delay is increased by approximately 5\%.  Threshold logic also has the advantage of absorbing CMOS logic so that the overall circuit design will be smaller.  An obfuscated cell library of 7 cells was designed and laid out for high level synthesis.  The library was used to obfuscate a Wallace Tree Multiplier and a FIR Filter.  Both circuits were compared to their nonobfuscated CMOS counterparts.  They showed approximately a 26\% decrease in area, 30\% decrease in dynamic power and 45\% decrease in leakage power than their static CMOS counterparts.

\section*{Acknowledgment}
We gratefully acknowledge NSF grants \#1237856 and \#1230401 for supporting this research.

\end{document}